\documentclass[12pt, a4paper]{article}
\usepackage{authblk}

\usepackage[latin1]{inputenc}
\usepackage[T1]{fontenc}
\usepackage{hyperref}

\usepackage{natbib} 
\usepackage{graphicx}

\usepackage{geometry}
\providecommand{\keywords}[1]{Keywords ---  \textit{#1}}
\providecommand{\orcid}[2]{#1 --- \textsc{#2} \\}

\begin{document}
\title{Spectroscopic Devices for Asteroseismology With Small Telescopes in NARIT}

\author{\textsc{Somsawat Rattanasoon}}
\author{\textsc{Eugene Semenko}\footnote{Corresponding author: eugene@narit.or.th}}
\author{\textsc{David Mkrtichian}}
\author{\textsc{Saran Poshyachinda}}
\affil{National Astronomical Research Institute of Thailand (Public Organization) \\ 260  Moo 4, T. Donkaew,  A. Maerim, Chiangmai, 50180 Thailand}
\date{}
\maketitle


%

\begin{abstract}
National Astronomical Research Institute of Thailand (NARIT) has a manifold network of small telescopes installed worldwide. These telescopes serve educational and research purposes and are equipped mainly with CCD detectors for direct imaging and photometry. To extend the possible field of applications, several telescopes were fitted with commercially available medium-resolution spectrographs eShel from Shelyak. With these devices, researchers in NARIT obtained a versatile tool for stellar spectroscopy. Here we describe the current status of available equipment, possible ways of upgrading, and briefly introduce the achieved results of the asteroseismologic study of fast-rotating stars.
\end{abstract}


\section{Motivation}
A fibre-fed medium-resolution echelle spectrograph eShel has been designed and distributed for small telescopes by Shelyak Instruments (France) since 2008 \citep{2011IAUS..272..282T}. A typical device consists of a stationary spectrograph block linked by a fibre with 50 $\mu$m core to the Fibre Injection and Guiding Unit (FIGU) installed at the telescope side. FIGU is also connected through a 200-$\mu$m fibre channel to the Calibration Unit comprising halogen, LED, and ThAr lamps. Spectrograph and its components are commercially available on the company's website \url{https://www.shelyak.com/}.
Earlier models of eShel registered spectra within the wavelength range 430--700\,nm with the resolution $R > 10,000$. In 2018, after the upgrade, which affected many components of eShel, the working range was significantly extended.

NARIT has a distributed network of small telescopes with apertures up to 1 m. For the spectroscopy of relatively bright stars, these telescopes can optionally be equipped with eShel. At the moment, NARIT has three devices with serial numbers 6H-115 (2010), 6H-128 (2016), and 6H-171 (2018). All spectrographs were acquired in their original complete set, thus having limited capabilities. To enable observations of fainter objects and to increase sensitivity in the blue part of the spectrum, we initiated a substantial upgrade of a device with SN 6H-171.

\section{Modification and Tests}
\begin{figure}
\centering
\includegraphics[width=12cm]{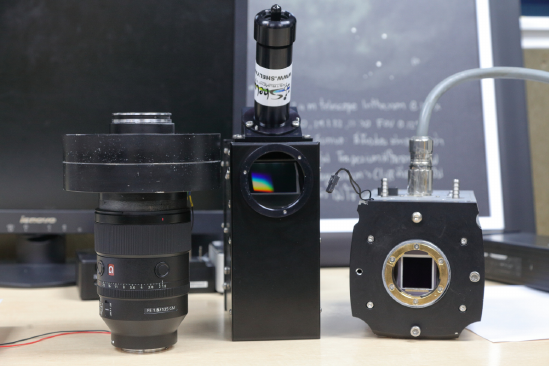}
\bigskip
\begin{minipage}{12cm}
\caption{Main elements of eShel upgraded in NARIT.}
\label{fig:figure1}
\end{minipage}
\end{figure}

The improved device received a new high-OH fibre with enhanced throughput in the blue part of the spectrum, a new doublet collimator (Shelyak provided both components), a new imaging lens, and a professional-grade CCD. All components, except fibre, are shown in Fig. \ref{fig:figure1}.
As a detector, we use a water-cooled Andor iKon-L system based on a $2048\times2048$ pixels CCD array with 13.5\,$\mu$m pixel pitch. To match the plate scale to the increased pixel size, among several lenses with comparable focus lengths available in the market, we choose a commercial lens Sony FE 135\,mm F1.8 GM, primarily due to its outstanding optical quality. Subsequent testing of the whole assembly also showed excellent transmission of the selected lens within the required range of wavelengths. The imaging lens is attached to the CCD camera through a specially designed adapter with an enclosed shutter.

Technical parameters of the original and upgraded versions of eShel are summarized in Table \ref{tab:summary}.

\begin{table}
\centering
\begin{minipage}{120mm}
\caption{Technical parameters of upgraded eShel. Asterisk indicates the new elements with the same listed characteristics.}
\label{tab:summary}
\end{minipage}
\bigskip

\begin{tabular}{l l l}
\hline
\textbf{Parameter} & \textbf{Original Value}  & \textbf{New value} \\
\hline
FIGU $F\#$       &   F6  & original  \\
Fibre core      &   50 $\mu$m & $^{*}$50 $\mu$m  \\
$f_\mathrm{col}$, $F\#_\mathrm{col}$       & 125 mm, F5 &  $^{*}$125 mm, F5 \\
$d_\mathrm{echelle}$, $\theta_\mathrm{b}$ & 79 mm$^{-1}$, 63$^\circ$ & original  \\
Absolute orders $\#$   &  32--52 & 24--57 \\
Imaging lens          &  Canon EF 85 F1.8  & Sony FE 135\,mm F1.8 GM \\
Detector (sensor)     &  ATIK 460 EX (Sony ICX694)   &  Andor iKon-L (E2V CCD42-40)  \\
Sensor format         &  $2749\times2199@4.54\mu$m   &  $2048\times2048@13.5\mu$m   \\
\hline
\end{tabular}
\end{table}

An upgraded variant of the spectrograph was installed for tests in a spectrograph room of the Thai National Observatory (TNO) at Doi Inthanon (Chiang Mai, Thailand) in a temperature-controlled environment. The FIGU was mounted to the left Nasmyth port of the 1-m telescope of TNO. Tests were performed in December 2022 and January 2023 under affordable weather conditions and were aimed at the verification of the optical performance of the assembly. Observational data include a standard set of calibrations (bias, flat, ThAr) and spectra of the selected stars and daytime sky. Two-dimensional raw FITS images were reduced using the pipeline PyYAP (\url{http://github.com/ich-heisse-eugene/PyYAP}), specially adapted to a new device.

\section{Results}

Test images taken with the upgraded device showed remarkable aberrations arising from the misaligned optical elements of the spectrograph. As this problem appeared in the direction perpendicular to dispersion, it influenced the overall throughput and the level of scattered light. Still, it didn't affect the spectral resolution and transmission of the device. Thus we leave the evaluation of the total throughput and stability for future works and concentrate here primarily on studying these unaffected characteristics.

\subsection{Transmission}
\begin{figure}
\centering
\includegraphics[width=14cm]{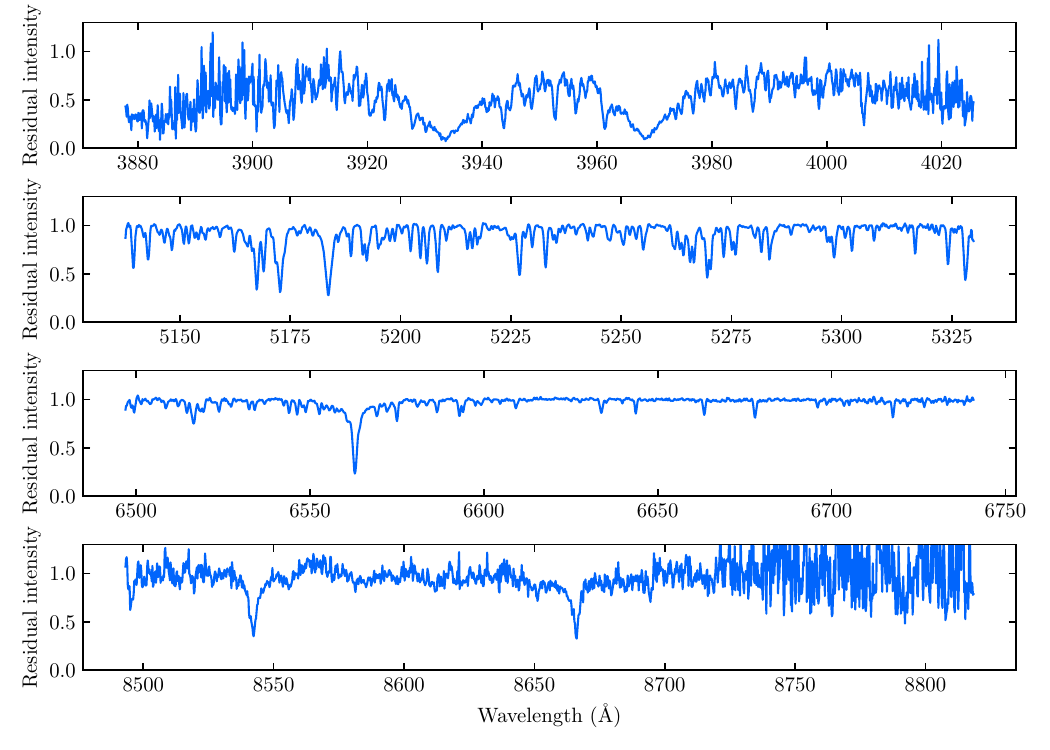}
\bigskip
\begin{minipage}{14cm}
\caption{Four samples of the solar spectrum taken at TNO during day time.}
\label{fig:figure2}
\end{minipage}
\end{figure}

Analysis of observational data revealed significantly improved spectrum quality due to better control of aberrations and enhanced transmission in the Sony lens. In the images, the point spread function remains nearly stable across the field of view in the 380-850 nm wavelength range. As a result, the shortwave limit of the working spectral range has been extended by 70 nm, from 450 nm to 380 nm. In the infrared, the working range of the current setup is limited by 900 nm. In Fig. \ref{fig:figure2}, we show four samples of the observed spectrum of the daytime sky.

\subsection{Resolving power}
The resolving power of the modified eShel was evaluated by fitting the Gaussian function to the emission lines of the ThAr spectrum. This procedure is implemented as a standard step of processing in PyYAP.

Inspection of the ThAr spectra showed that the focus of the imaging camera remained stable during all observational nights. Within the spectrograph's working wavelength range, the resolving power $R = \lambda/\Delta\lambda$ varied from $10,000$ to $12,500$, with the median $R = 11,700$ evaluated from 355 lines in a single image. The resolving power does not variate significantly between nights: the full width at half maximum (FHWM) of the mean ThAr line equals 3.7 pixels, close to the optimal sampling.

\section{Scientific Application}
A medium-resolution fibre-fed spectrograph, in combination with a 1-meter class telescope, can be a powerful instrument for the spectroscopy of relatively bright sources. Literature has many examples of using eShel in stellar physics and the physics of the Solar system objects. Due to its compact design and high positional stability, this spectrograph appears even in the observations of the extrasolar planets. The thing is that the accuracy of the radial velocity measurements reported in \cite{2014CoSka..43..451K},  \cite{2015AN....336..682P}, and \cite{2017PASP..129f5002E} was better than 100\,m\,s$^{-1}$ for the stars brighter than 11 magnitudes and exposure time under one hour. Such characteristics enable the detection and observation of hot Jupiters around the brightest stars. \cite{2014CoSka..43..451K} also gave the example of how to use eShel for observation of pulsating stars (cepheids). 

The proposed upgrade opens new perspectives for the family of small telescopes in NARIT, as we have several spectrographs which, after the improvement, can be installed at any of our telescopes. In this way, it becomes possible to move part of scientific proposals aimed at studying exoplanets, active solar-like stars, binary and multiple stars from the main 2.4-m Thai National Telescope to smaller instruments without losing the efficiency and observing time. However, the main stimulus which led us to this technical work was the capability of using this device for asteroseismology of the brightest fast-rotating pulsating stars.

\begin{figure}
\centering
\includegraphics[width=15cm]{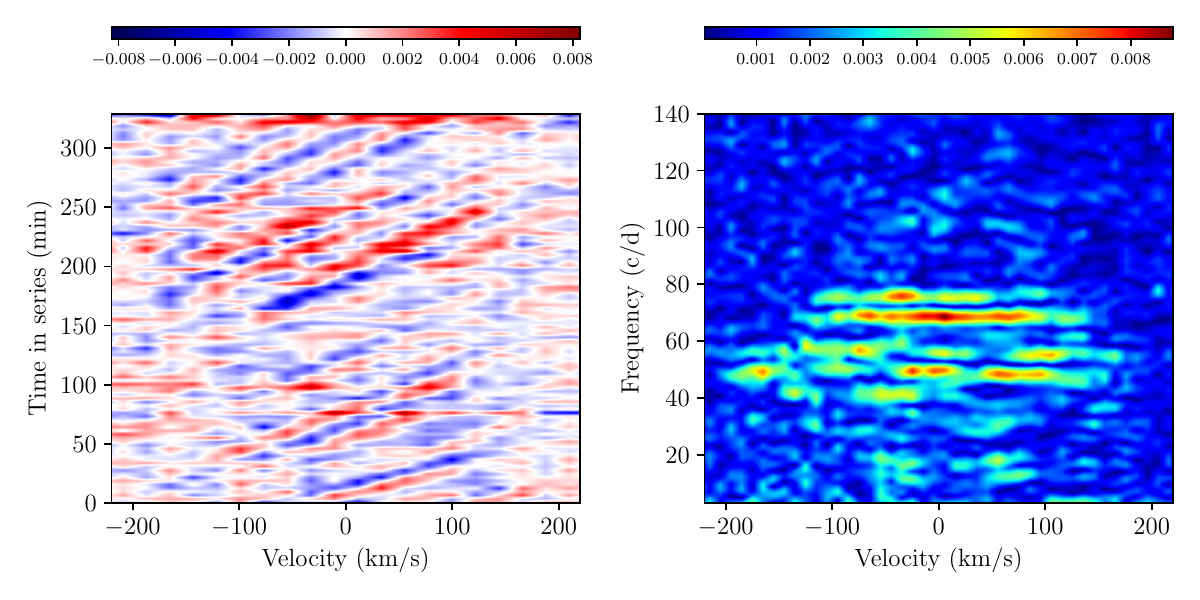}
\bigskip
\begin{minipage}{14cm}
\caption{Dynamic spectrum and a 2D periodogram used for the discovery of non-radial pulsation in a bright fast-rotating A-type star.}
\label{fig:figure3}
\end{minipage}
\end{figure}

To demonstrate the efficiency of eShel in asteroseismological observations, in Fig. \ref{fig:figure3}, we show an example of non-radial pulsations discovered in a 4-magnitude fast-rotating star. A typical pattern of waves propagating across the averaged spectral profile is in the left panel of \ref{fig:figure3}. The right panel shows the 2D periodogram used for the identification of frequencies of pulsation. In this example, the star has been observed continuously with short exposures for more than five hours with the original version of eShel and the 1-m telescope of NARIT. The upgraded version of the spectrograph will allow us to increase the signal-to-noise ratio (SNR) of observational data and, thus, expand the number of potential targets or increase the temporal resolution of data with shorter exposure time preserving the same level of SNR.

\subsubsection*{ORCID identifiers of the authors}
\orcid{0000-0002-1912-1342}{Eugene Semenko}
\orcid{0000-0001-5094-3910}{David Mkrtichian}

\subsubsection*{Author contributions}
SR, ES, and DM are responsible for formulating the project, its technical implementation, and carrying out the observations. ES and DM are responsible for data reduction and analysis. SP contributed to the project administration. All authors equally contributed to the text of the article.

\subsubsection*{Conflicts of interest}
The authors declare no conflict of interest.

\bibliographystyle{bullsrsl-en}
\bibliography{extra}

\end{document}